\documentstyle[12pt,aasms4]{article}
\input epsf.sty
\newcommand{\groj}{\mbox{GRO~J1655$-$40}}
\newcommand{\frank}{\mbox{GRO~J1719$-$24}}
\received{1995 Oct.\ 16}
\accepted{1996 Feb.\ 19}
\begin{document}
\title{Search For Rapid X-ray Variability from the Black-Hole 
Candidate GRO~J1655-40}
\author{D. J. Crary\altaffilmark{1}, C. Kouveliotou\altaffilmark{2}, 
        J. van Paradijs\altaffilmark{3,4}, 
        F. van der Hooft\altaffilmark{4}, M. van der Klis\altaffilmark{4},
        \linebreak
        B. C. Rubin\altaffilmark{2}, D. M. Scott\altaffilmark{2},
        M. H. Finger\altaffilmark{2}, and B. A. Harmon\altaffilmark{5}}
\altaffiltext{1}{NAS/NRC Research Associate, NASA Code ES-84, Marshall 
    Space Flight Center, Huntsville, AL\ \ 35812, USA}
\altaffiltext{2}{Universities Space Research Association, Huntsville,
    AL\ \ 35806, USA}
\altaffiltext{3}{Department of Physics, University of Alabama in Huntsville,
    Huntsville, AL\ \ 35899, USA}
\altaffiltext{4}{Astronomical Institute ``Anton Pannekoek'', University
    of Amsterdam \& Center for High-Energy Astrophysics, Kruislaan 403, 
    NL-1098 SJ Amsterdam, The Netherlands}
\altaffiltext{5}{NASA/Marshall Space Flight Center, Huntsville, AL 35812, USA}
\begin{abstract}
We have examined fifteen days of CGRO/BATSE data, obtained during the first 
outburst of the black-hole candidate source \groj, to search for rapid 
variability of its X-ray flux. We find no evidence for 
significant variability of \groj\ during our observations, with 
a 2$\sigma$ upper limit 
to the fractional r.m.s.\ amplitude in the frequency range 0.03--0.488 Hz 
of $6.6\%$. We cannot, on the basis of our 
observations, determine the source state (low, high, or very-high 
state) of \groj.  
\end{abstract}
\keywords{stars: individual (GRO~J1655$-$40) --- X-rays: stars}

\section{Introduction}

The X-ray transient source \groj\ (also known as 
X-ray Nova Scorpii 1994) was discovered
with BATSE on the Compton Gamma Ray Observatory on 27 July 1994.
The evolution of the hard (20--100 keV) X-ray intensity during the 
subsequent 5 months has been discussed by Harmon et al.\ (1995). 
Three outbursts were observed, each characterized by
a fast ($< 1$ day) rise to a level of 600--700 mCrab in the 20--100 keV range,
with no well-defined 
single maximum during the outbursts (10--50 days). 
Typical peak flux levels were $0.30\; 
\rm photons \; cm^{-2}\, sec^{-1}$ in the energy range of
the observations.
There was no significant emission from the location of \groj\
for a 50 day period prior to 27 July.

\groj\ is a nearby (2--3 kpc) dynamical black-hole
candidate with a reported mass function 
of $3.35 \pm 0.14\rm\, M_\odot$ (Bailyn et al.\ 1995a,b; see Tanaka \& 
Lewin 1995 for a recent review of black-hole candidates).  The 
source shows strong radio outbursts associated
with superluminal expansion events (Tingay et al. 1995; Hjellming \& 
Rupen 1995) that
are correlated with the increase in hard X-ray flux.
Bailyn et al.\ (1995b)
found that the optical light curve shows
eclipses, at an orbital period of 2.6 days, and suggested that
the orbital inclination of \groj\ is close to 90$^\circ$.
The energy spectrum can be well described
by a power law out to at least 300 keV, with spectral
(photon) index varying between 2.5 and 3.1 (Wilson et al. 1995).  

Here we report on the analysis of the fast variability of the hard X-ray 
intensity of \groj.  In section 2 we describe the observations
and data analysis.  We discuss our results in section 3. Section 4
summarizes our conclusions.

\section{Observations}

We used the 1.024 second time resolution count rates from the BATSE Large
Area Detectors (LADs) in 2 energy channels 
corresponding to the energy range 20--50 and \mbox{50--100 keV} 
(Fishman et al.\ 1991).  Intervals were
selected of 512 time bins (524.288 seconds) without interruptions or 
data drop-outs, during which \groj\ was above the Earth horizon.
Our chosen segment length is a compromise
between the number of contiguous segments
that could be obtained
(this is determined by incomplete data coverage, and SAA passages) and
frequency resolution.
We used data only from detectors in which the source was seen at an
angle of $< 60^\circ$ from the detector normal. 
The segments were quadratically detrended, the resulting time series
converted to power density spectra (PDS), and the PDS averaged
over an entire day. 
Approximately 45 intervals were obtained
each day during the outbursts.
We have determined that for the purposes of this analysis quadratic
detrending is equivalent to a detailed background model subtraction
as described by Rubin et al.\ (1993).
Using similar intervals obtained when the source
was occulted by the Earth (approximately 29 per day) we verified
that the PDS of the detector background counting rate is flat down to 
at least 0.03 Hz.

The data set that we could use is constrained by the presence of 
other potential noise
sources in the field of view of the uncollimated LAD detectors that are facing
\groj.  Cygnus~X-1 is a persistent source that shows strong variability in the
frequency interval of interest.  Because of the large angle ($\sim\!
86^\circ$) between Cygnus~X-1 and \groj, it is possible to choose detectors
that face \groj\ in which Cygnus~X-1 is at an angle $> 90^\circ$ from the
detector normal. Only these detectors have been used in the following
analysis. The soft X-ray transient source 
\mbox{GX~339$-$4}
(located within $10^\circ$ of \groj) was found from the source
monitoring by BATSE through daily 
occultation analysis (Harmon et al.\ 1993) to be bright 
during the second and third outburst of \groj.
We have therefore confined our observations to the 15 days corresponding
to the first outburst when no other sources in the vicinity of 
\groj\ were detected with BATSE at a level higher than 
$0.16 \rm \; photons \; cm^{-2}\, sec^{-1}$, and when	
the daily averaged flux of \groj\ was higher than
$0.19\; \rm photons\; 
cm^{-2}\, sec^{-1}$ (28 July 1994--11 August 1994).  
Figure~1 shows the flux levels for this
interval (Harmon et al.\ 1995).

The analysis is complicated by the existence of background noise
in the LAD detectors that is likely due to weak unresolved sources in the
uncollimated field of view.  Figure 2 shows the power levels 
normalized to a Poisson level of 2.0 (Leahy et al.\ 1983),
and averaged
over the frequency interval \mbox{0.03--0.488 Hz} in
detectors facing the region of \groj\ for 16 days prior to the
outburst.  During this period, no transient sources with a hard spectrum 
and known fast variability were detected with BATSE in this
region
having a flux greater than $0.2 \rm\; photons\; cm^{-2}\, sec^{-1}$ 
(20--100 keV). 
The 16 day average of the (0.03--0.488 Hz) 
power density is $2.035 \pm 0.007$. 

Figure 3 shows the averaged PDS for \groj\ during the interval 28 July 1994
to 11 August 1994, corrected for the background 
noise effect described above.
The power density was normalized to represent the squared
fractional r.m.s.\ amplitude per unit frequency 
(see Van der Klis 1995 for the formulae).
Figure 4 shows the (0.03--0.488 Hz) integrated daily averaged PDS 
measured from \groj, where the PDS have been normalized in the same way as 
in Figure 3.
The errors have been 
propagated from the variance per bin calculated for each daily average,
the errors in the subtracted background power, and the count rate errors
derived from occultation analysis. 

The average value of this squared fractional r.m.s. amplitude equals $(0.6
\pm 1.9) \times 10^{-4}$ (r.m.s./mean)$^2$; the $2\sigma$ 
upper limit to this quantity equals $4.4 \times 10^{-4}$. The corresponding 
$2\sigma$ upper limit to the fractional r.m.s. amplitude, $f$, averaged 
over the 15 days of observation equals $6.6 \%$.

\section{Discussion}

Based on their spectral and fast-variability properties, it appears that the 
black-hole X-ray binaries have three states, the \lq low', \lq high' 
and \lq very 
high' states (Miyamoto et al.\ 1994; Van der Klis 1994a). The high state is 
characterized by the presence of a strong ultrasoft component which 
dominates the 1--10 keV X-ray flux. Since no low-energy X-ray observations 
were made during the period covered by our study, we cannot make a 
statement on the source state of \groj\  on the basis of its spectral 
characteristics.

In the low state the X-ray spectra of black-hole X-ray binaries are very 
hard, and well described by a power law that 
extends up to several hundreds of keV. The PDS is characterized by a power 
law at high frequencies, with an approximately flat part below a variable 
\lq break frequency' $\nu_{\rm b}$. In PDS that are normalized according to 
Belloni \& Hasinger (1990) the high-frequency 
part of the PDS does not change much; therefore, $f$, 
in a given frequency interval that contains the break 
frequency, is anticorrelated with $\nu_{\rm b}$.

If \groj\ had been in the low state during our observations, 
and followed the ($\nu_{\rm b},f$) relation of Cygnus~X-1 (Van der Klis 
1994b; Crary et al. 1995) the break frequency would have to be at 
least $\sim 0.7$ Hz, to account for the observed upper limit to $f$ of 6.6\%. 
The required extrapolation of $\nu_{\rm b}$ beyond the range observed for 
Cygnus~X-1 (0.04 to 0.4 Hz) seems moderate enough that we cannot exclude that 
\groj\ was in the low state during our observations.

The high state of black-hole X-ray binaries is characterized by the 
presence of a strong ultra-soft spectral component, which dominates the 
flux in the 1--10 keV range. In the high state an anticorrelation is 
observed between the 2--10 keV source variability and spectral hardness, 
which has been interpreted as a dilution of the variability in the hard 
power law spectral component by the less variable ultrasoft component.

No information on the presence or absence of an ultrasoft component 
in the X-ray spectrum of \groj\ during our observation is available. 
If we assume that the total luminosity of \groj\ does not exceed 
the Eddington limit, then with the distance of 3.5 kpc (Bailyn et al. 1995a) 
and mass of at least 3.35 M$_{\odot}$ we find that there is ample room for 
an ultrasoft component (the 2--100 keV luminosity in the power law component 
is only one third of the Eddington limit).
We cannot, therefore, exclude that \groj\ was then in the high state. 
If it were, the low value of $f$ we have observed would indicate that the
variability of the hard (20-100 keV) power law spectral component is smaller
in the high state than in the low state.  This, in turn, would suggest
that the low values of $f$ observed in the 2-10 keV band in the high
state is not only the result of the above mentioned dilution
by the ultrasoft component, but that the variability of the power
law component in this energy range is low as well. 

The very high state (VHS) has so far only been observed in two sources, 
GS~1124$-$68 (Miyamoto et al.\ 1993) and GX~339$-$4 (Miyamoto et al.\ 1992). 
The X-ray spectrum then has, at least part of the time, a power law 
component above 20 keV. 
Miyamoto et al.\ (1994) have decomposed the PDS obtained with 
Ginga for GS~1124$-$68 in the VHS into a power law component and 
a \lq flat-topped' component. The latter has a constant value of 
$\sim 0.001\; {\rm Hz}^{-1}$ for frequencies below 1 Hz, and is independent 
of energy in the 1--37 keV range. This PDS level corresponds to a value 
for $f$ in the frequency range 0.03--0.488 Hz of 3\%. If we assume that 
the energy independence of the PDS in the VHS extends throughout the BATSE 
energy range, the observed variability properties of \groj, 
presented here are consistent with the idea that the source was in the VHS 
during our observations. 

\section{Conclusions}

During the first outburst of \groj\ we detected no (0.03--0.488 Hz) 
variability in its (20--100 keV) flux, with an upper limit to its r.m.s. 
amplitude of 6.6\% ($2 \sigma$). 
Several bright ($\geq 1$ Crab, 20--100 keV)  black-hole candidate sources 
observed with BATSE (in particular, Cygnus X-1, \frank, 
and GRO~J0422+32) have shown enhanced noise power in the 
0.03--0.488 Hz range, 
and occasional quasi-periodic oscillations (Kouveliotou 1994; Van der Hooft 
et al. 1995).
Fractional r.m.s.\ values for this 
noise observed from Cygnus~X-1 (10--30\%, Crary et al. 1995)
and \frank\ ($\sim \rm 15\%$, Van der Hooft et al. 1995) are 
characteristic of the low state (Van der Klis 1994), and it is likely that 
they were encountered in that state with BATSE. 

Based on the BATSE data for the period July 28--August 11, 1994 alone we 
cannot interpret the observed low variability of \groj\ in terms of 
the source state scheme for black-hole candidates. For that a better 
understanding of the hard X-ray properties of black-hole candidates (e.g., 
their spectral slope) in the different source states is required.
Investigations of this type will be facilitated by 
low-energy ($< 10$ keV) observations, performed 
concurrently with a BATSE observation. 

\vspace{.25in}
We thank Dr. W. Lewin for his comments on this paper. 
This project was performed within NASA grant NAG5-2560 and supported
in part by the Netherlands Organization for Scientific Research (NWO)
under grant \mbox{PGS 78-277.} This work was performed while DJC held
a National Research Council-NASA Research Associateship.
FvdH acknowledges support by the Netherlands Foundation for Research in
Astronomy with financial aid from NWO under contract number
\mbox{782-376-011}, and
the Leids Kerkhoven--Bosscha Fonds for a travel grant.
JvP acknowledges support from NASA grant \mbox{NAG5-2755}.

\pagebreak

\pagebreak
\noindent Fig. 1 -- Hard X-ray light curve (20-100 keV) for GRO~J1655$-$40 
from TJD (= Julian Day $-$ 2440000.5) 9560 
(1995 July 27) to 9585 (1995 August 21) obtained with BATSE occultation
analysis.

\noindent Fig. 2 -- Average power per bin for detectors facing the region of 
\groj\ for 16 days prior to discovery.  The power has
been normalized such the Poisson level
is 2.0.  The average taken over the interval 0.03--0.488 Hz is 
$2.035 \pm 0.007$, shown by the dotted line.

\noindent Fig. 3 -- Averaged (1995 July 28--1995 August 11) PDS corrected
for Poission fluctuations and background noise shown in Figure 2,
normalized to fractional squared r.m.s.\ Hz$^{-1}$.

\noindent Fig. 4 -- Calculated fractional squared r.m.s.\ 
(0.03--0.488 Hz) for \groj\ during
outburst from TJD 9561 (1995 July 28) to 9575 (1995 Aug 11). Contributions
to the background shown in Figure 2 have been subtracted from these data.

\pagebreak
\pagestyle{empty}
\begin{figure}
\epsfysize=350pt
\vspace{-1in}
\epsffile{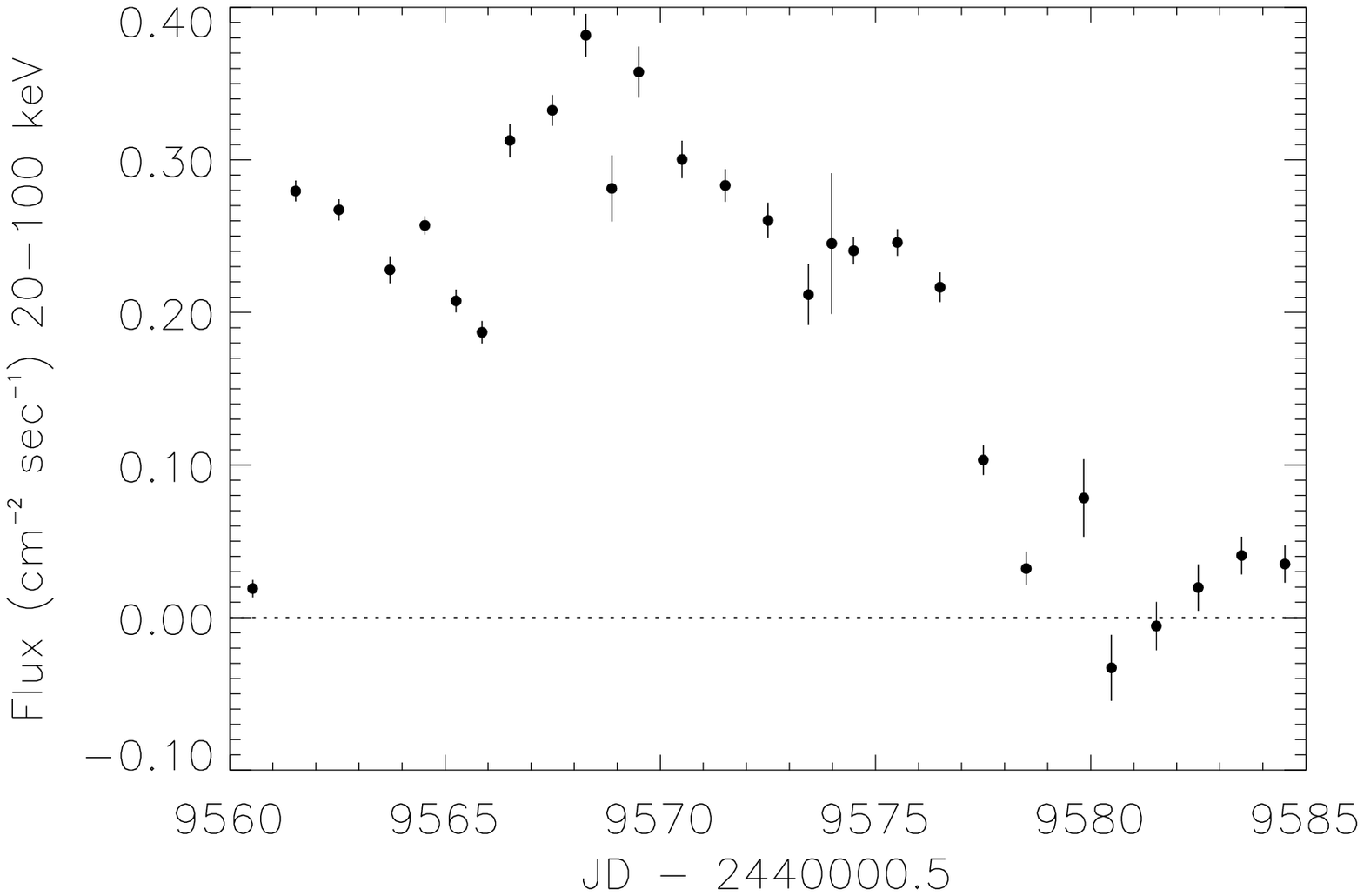}
\caption{}
\end{figure}

\begin{figure}
\epsfysize=350pt
\vspace{-1in}
\epsffile{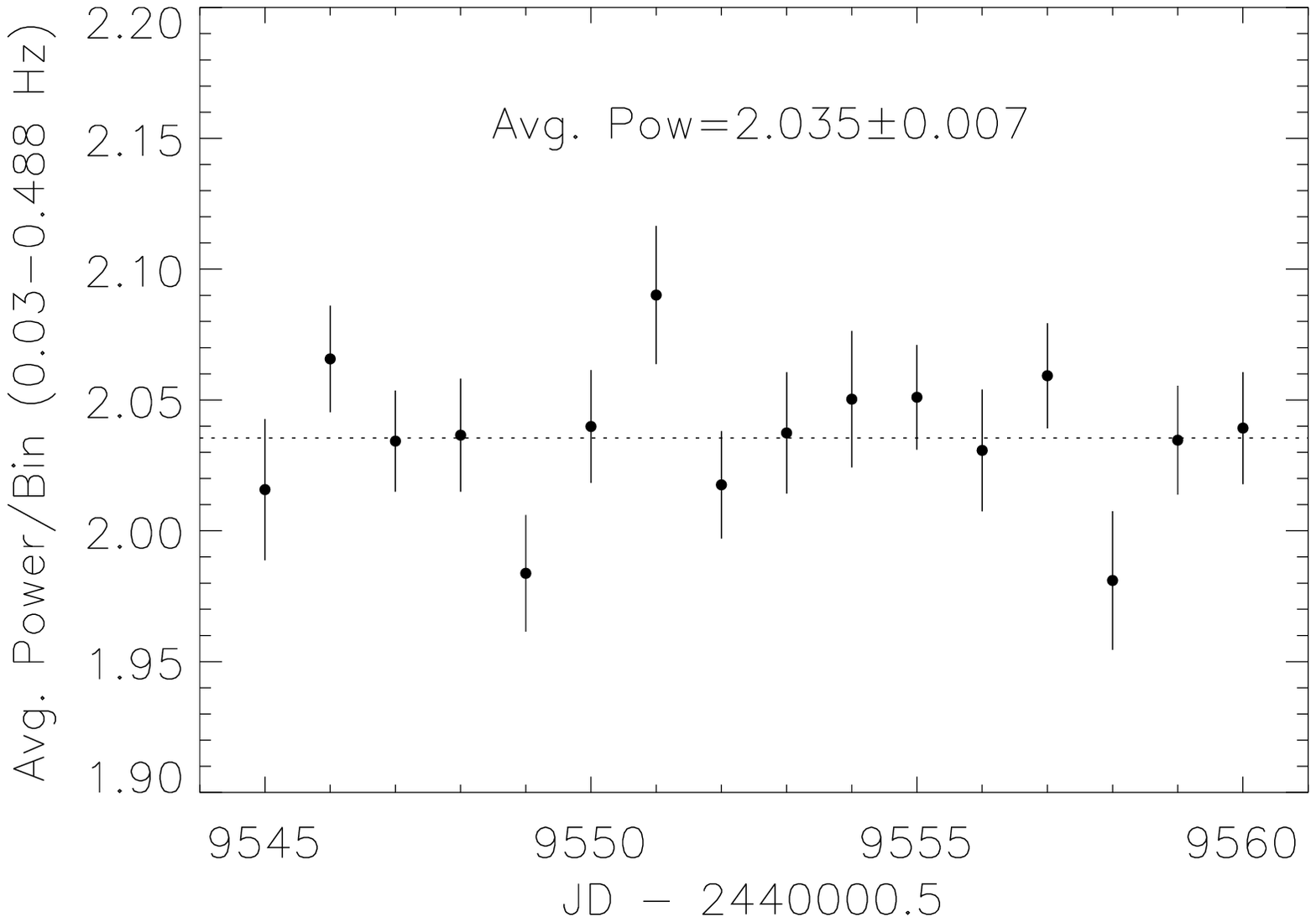}
\caption{}
\end{figure}

\begin{figure}
\epsfysize=350pt
\vspace{-1in}
\epsffile{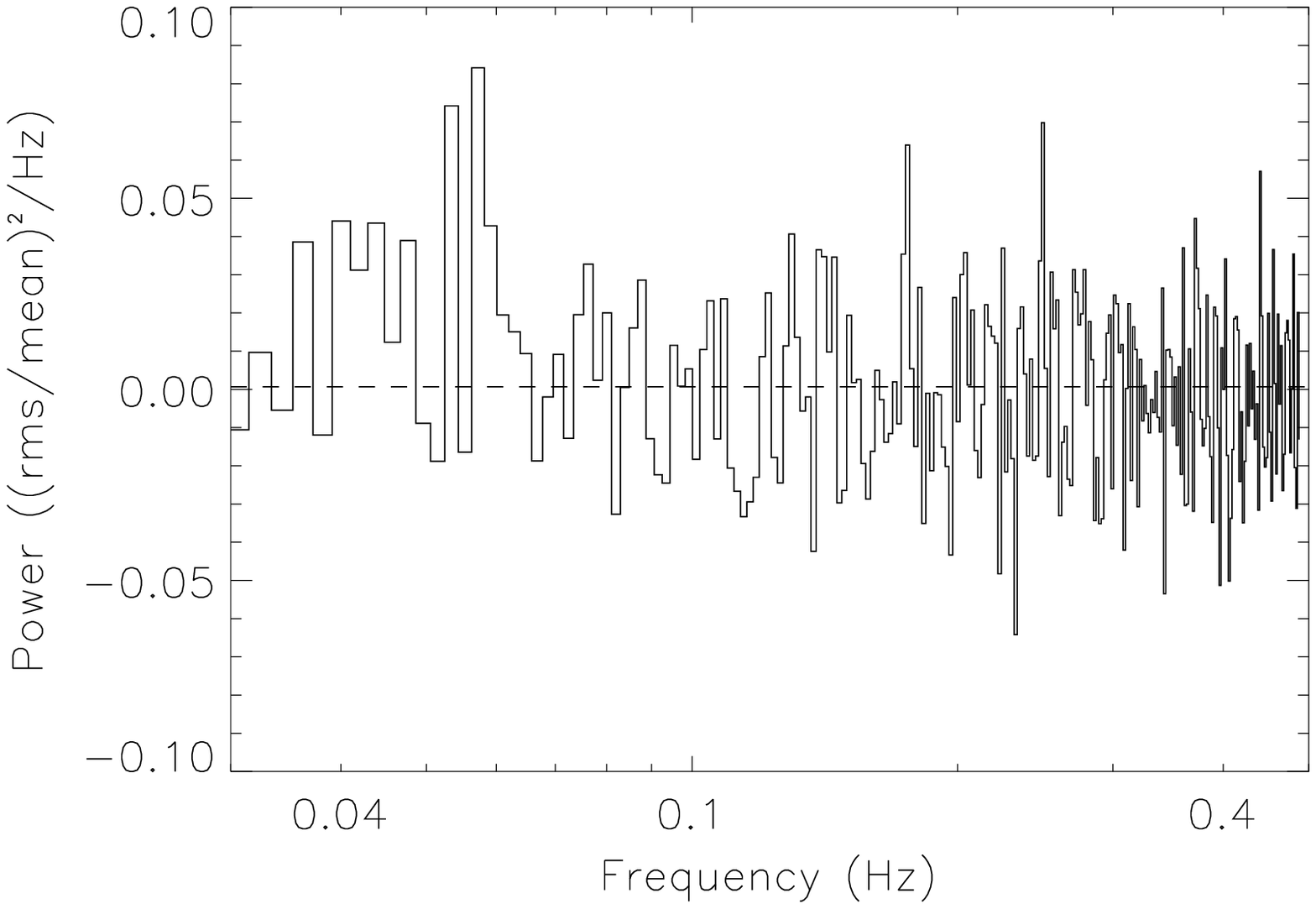}
\caption{}
\end{figure}

\begin{figure}
\epsfysize=350pt
\vspace{-1in}
\epsffile{rms22.ps}
\caption{}
\end{figure}

\end{document}